\documentstyle[11pt,aaspp4,flushrt]{article}

\begin{document}

\title{ON THE MASS OF POPULATION III STARS}
\author{Fumitaka Nakamura} 
\affil{Faculty of Education and Human Sciences, Niigata University,
Ikarashi 2-8050, Niigata 950-2181, Japan}
\authoremail{fnakamur@ed.niigata-u.ac.jp} 

\and

\author{Masayuki Umemura}
\affil{Center for Computational Physics, University of Tsukuba,
Tsukuba, Ibaraki 305-8577, Japan}
\authoremail{umemura@rccp.tsukuba.ac.jp}

\begin{abstract}

Performing one-dimensional hydrodynamical calculations coupled with 
non\-equilibrium processes for hydrogen molecule formation, we pursue
 the thermal and dynamical evolution of filamentary primordial gas 
clouds and attempt to make an estimate on the mass of population III stars. 
The cloud evolution is computed from the central proton density 
$n_c \sim 10^{2-4}$ cm$^{-3}$ 
up to $\sim 10^{13}$ cm$^{-3}$. It is found that, almost independent 
of initial conditions, a filamentary 
cloud continues to collapse nearly isothermally due to H$_2$ cooling 
until the cloud becomes optically 
thick against the H$_2$ lines ($n_c \sim 10^{10-11}$ cm$^{-3}$). 
During the collapse the cloud structure separates into two parts, 
i.e., a denser spindle 
and a diffuse envelope. The spindle contracts quasi-statically, and 
thus the line mass 
of the spindle keeps a characteristic value determined solely by the 
temperature ($\sim 800$ K), 
which is $\sim 1 \times 10^3M_\odot$ pc$^{-1}$ during the contraction
 from $n_c \sim 10^{5}$ 
cm$^{-3}$ to $10^{13}$ cm$^{-3}$. Applying a linear theory, we find 
that the spindle is unstable
 against fragmentation during the collapse. The wavelength of the 
fastest growing perturbation ($\lambda _{\rm m}$) 
lessens as the collapse proceeds. Consequently, successive fragmentation 
could occur. When the central density exceeds $n _c \, \sim \, 10^{10-11}$ 
cm$^{-3}$, the successive fragmentation may cease since the cloud becomes 
opaque against the H$_2$ lines and the collapse decelerates appreciably.
Resultingly, the minimum value of $\lambda _{\rm m}$ is estimated to be 
$\sim 2 \times 10^{-3}$ pc. The mass of the first star is then expected 
to be typically $\sim 3 M_\odot$, which may grow up to $\sim 16 M_\odot $
 by accreting  the diffuse envelope. Thus, the first-generation stars are 
anticipated to be massive but not supermassive.

\end{abstract}


\section{Introduction}

Based on the standard Big Bang nucleosynthesis, 
the first generation of stars should form from materials deficient 
in heavy elements. In the present-day galaxies, heavy elements or dust grains 
provide the most efficient cooling mechanism, while
the cooling process in primordial gas is likely
to be governed by hydrogen molecules.
Many authors hitherto have considered the formation processes of
primordial stars from metal deficient gas
(e.g., Matsuda, Sato, \& Takeda 1969; Yoneyama 1972; 
Hutchins 1976; Silk 1977a, 1977b;
Yoshii \& Sabano 1979;  Carlberg 1981; 
Struck-Marcell 1982a, 1982b; Lepp \& Shull 1983, 1984; 
Silk 1983; Palla, Salpeter, \& Stahler 1983; 
Yoshii \& Saio 1986; Shapiro \& Kang 1987;
de Ara\'{u}jo \& Opher 1989; Uehara et al. 1996; Haiman, Thoul, \& Loeb 1996;
Omukai et al. 1998).
Such first-generation stars, say Population III, 
could play an important role in the early evolution of galaxies
(e.g., Tegmark, Silk, \& Blanchard 1994; 
Ostriker \& Gnedin 1996) or the formation of massive black holes 
(Umemura et al. 1993).
Also, they may be responsible for the chemical pollution
of intergalactic medium which is recently inferred from metallic absorption
in Ly$\alpha$ forest seen in quasar light 
(Cowie et al. 1995; Songaila \& Cowie 1996).
It is thus important to study 
the thermal and dynamical evolution of the primordial gas
clouds from which the first-generation stars would form.

If the hydrogen gas were to remain purely atomic, 
the primordial gas would be cooled down to 10$^4$ K 
due to the Lyman $\alpha$ lines.
It is, however, difficult for the gas temperature 
to become lower,  because hydrogen atoms
are poor radiator in the lower temperature gas.
Therefore, the Jeans mass of such a cloud,
which is often referred to the characteristic mass
in the star formation theory, becomes much greater than 
a typical stellar mass. 
In practice, hydrogen molecules provide key cooling mechanisms.
In contrast to the molecule formation on dust grains in
metal-rich interstellar gas, the formation of primordial
hydrogen molecules can proceed through the gas phase
reaction (Saslaw \& Zipoy 1967; Peebles \&Dicke 1968),
\begin{equation}
e \, + \, {\rm H} \, \rightarrow \, {\rm H}^{-} \, + \, h \nu
\label{eq:hm1}
\end{equation}
\begin{equation}
{\rm H}^{-} \, + \, {\rm H} \, \rightarrow \, {\rm H} _2 \, 
+ \, e
\label{eq:hm2}
\end{equation}
and 
\begin{equation}
{\rm H}^+ \, + \, {\rm H} \, \rightarrow \, {\rm H}_2 ^+ \, + \, h \nu
\end{equation}
\begin{equation}
{\rm H}_2 ^+ \, + \, {\rm H} \, \rightarrow \, {\rm H} _2 \, 
+ \, {\rm H}^+ .
\end{equation}
At high density of $n \, \gtrsim 10^8 $cm$^{-3}$, 
the molecules can also form through the three-body
reactions (Palla et al. 1983),
\begin{equation}
3 {\rm H}  \, \rightarrow {\rm H}_2 \, + \, {\rm H}
\label{eq:3h process1}
\end{equation}
and
\begin{equation}
2 {\rm H} \, + \, {\rm H}_2 \, \rightarrow \, 2 {\rm H}_2 .
\label{eq:3h process2}
\end{equation}
Even if a relatively small fraction ($\sim10^{-3}$) of the molecules form,
they make a significant contribution to cooling
through the rotational and vibrational transitions, so that
the temperature of primordial gas can be reduced to a lower temperature
than $10^4$K (e.g., Matsuda et al. 1969; Yoneyama 1972; 
Hutchins 1976; Palla, et al. 1983).
Resultantly, the Jeans mass could decrease to stellar mass.
However, although a lot of elaborate analyses have been made so far, 
the masses of Population III stars have not been well converged. 
Carlberg (1981) and Palla et al. (1983) have shown 
the Jeans mass can go down to a level of $\lesssim 0.1M_\odot$. 
Yoshii \& Saio (1986) derive the initial mass function (IMF) based upon the
opacity-limited fragmentation theory (Silk 1977a, 1977b) and find the peak of the IMF
around $4-10M_\odot$.  Uehara et al. (1996) have shown that the minimum
masses of the first-generation stars are basically determined by the
{\it Chandrasekhar} mass, say $\sim 1M_\odot$ (see also Rees 1976). 
In this paper, we reanalyze the formation of Population III stars
by computing the collapse of filamentary clouds
coupled with H$_2$ formation.

In the bottom-up scenarios like a cold dark matter (CDM) model, 
it is expected that the overdense regions with the masses 
of $10^5 - 10^7 M_\odot$ would
first collapse at the redshift range of $10 \lesssim z \lesssim 100 $ and
the first generation of stars would form there.
The importance of the H$_2$ cooling in the collapse of 
such primordial clouds has been stressed by several authors
(e.g., de Ara\'{u}jo \& Opher 1989; Haiman et al. 1996; 
Susa et al. 1996;  Tegmark et al. 1997).
It has been found in these studies that the mass fraction
of H$_2$ can reach $10^{-4}$ up to $10^{-3}$  and 
the temperature of the cloud can be reduced to $10^2- 10^3$ K.
However, most of the studies are restricted to highly simplified models
such as homogeneous pressure-less and/or spherical collapses.
In practice, the cloud contraction proceeds inhomogeneously.
Since the cloud is more or less nonspherical,
the deviation from spherical symmetry grows in time due to
its self-gravity until a shocked pancake forms (e.g., Umemura 1993).
Recently, several authors have studied the collapse of 
pregalactic clouds with multi-dimensional hydrodynamical simulations
(e.g., Anninos \& Norman 1996; Ostriker \& Gnedin 1996).
Unfortunately they still do not have enough spatial resolution
as well as mass resolution to explore the star formation.

The pancake could be gravitationally unstable 
against fragmentation (e.g., Umemura 1993; Anninos \& Norman 1996).
It tends to fragment into thin filamentary clouds rather than
spherical ones (Miyama, Narita, \& Hayashi 1987a, 1987b; 
Uehara et al. 1996).
The filamentary cloud can also fragment into smaller and denser cores
(e.g., Larson 1985), in which consequently stars can form.
In this paper we thus investigate the thermal and dynamical
evolution of the filamentary primordial gas clouds.
With using one-dimensional axisymmetric hydrodynamical scheme,
we pursue the evolution in the range of more than eighth order of
magnitude in density contrast to properly estimate the mass scale of 
the first-generation stars.

In \S \ref{sec:model construction}, we describe our model clouds
and the computational methods.
Numerical results are presented in \S \ref{sec:results}.
In \S \ref{sec:fragmentation} 
we consider the fragmentation of a collapsing filamentary cloud
and estimate the mass scale of the fragments and thereby
the mass of the first-generation stars.
\S \ref{sec:sec6} is devoted to the conclusions.

\section{MODEL}
\label{sec:model construction}

\subsection{Basic Equations}

To pursue the thermal and dynamical evolution of a 
filamentary primordial cloud, 
we employ a one-dimensional hydrodynamical
scheme.  We assume that the system is axisymmetric and 
that the medium consists of ideal gas.
The adiabatic index, $\gamma$, is taken to be 5/3 for
monatomic gas and 7/5 for diatomic gas.
We deal with the following nine species:
e, H, H$^+$, H$^-$, H$_2$, H$_2 ^+$, He, He$^+$, and He$^{++}$, 
The abundance of helium atoms is taken to be 10 \% of
that of hydrogen by number. 

The basic equations are then given in the cylindrical coordinates 
$(r, \varphi, z) $ by
\begin{equation}
{\partial \rho \over \partial t} \, + \, {1 \over r} {\partial \over \partial r
}
( \rho v_r) \, = \, 0 \; ,
\label{basic eq. 1}
\end{equation}
\begin{equation}
{\partial \over \partial t} (\rho v_r ) 
\, + \, {1 \over r} {\partial \over \partial r} (r \rho v_r ^2 ) 
\, + \, {\partial P \over \partial r} \, + \, 
\rho {\partial \psi \over \partial r}  \, = \, 0 \; ,
\label{basic eq. 2}
\end{equation}
\begin{equation}
{\partial E \over \partial t} 
\, + \, {1 \over r} {\partial \over \partial r} [ r v_r (E \, + \, 
P)] \, + \, \rho {\partial \psi \over \partial r} v_r \, + \, 
\Lambda _{\rm net} \, = \, 0 \; ,
\label{basic eq. 3}
\end{equation}
\begin{equation}
\triangle \psi \, = \, 4 \pi G \rho \; ,
\label{basic eq. 4}
\end{equation}
where
\begin{equation}
\rho \, = \, \sum _i \rho _i \; ,
\end{equation}
\begin{equation}
P \, = \, \sum _i P_i \, = \, \sum _i n_i k T \; ,
\label{basic eq. 5}
\end{equation} 
\begin{equation}
E \, = \, \sum _i \left({P_i \over \gamma _i \, - \, 1} \, + \, {1 \over 2}
\rho _i v_r ^2 \right) \; ,
\label{basic eq. 6}
\end{equation}
where $\rho$, $n$, $v_r$, $P$, $\psi$, $E$, 
$G$, and $k$ are the mass density, number
density, radial velocity, gas pressure, gravitational
potential, energy per unit volume, 
gravitational constant, and Boltzmann constant, respectively.
The symbol $\Lambda _{\rm net} $ denotes the
cooling function which represents the net energy loss rate  per unit volume.
The values with subscript $i$ denote those of the $i$-th species.

The number density of the $i$-th species, $n_i$, is obtained 
by solving the following time-dependent rate equations,
\begin{equation}
{d x_i \over dt} \, = \, n_H \sum ^9 _{j = 1} \sum ^9 _{k = 1} k_{jk} x_j 
x_k
\, + \, n_H ^2 \sum ^9 _{l = 1} \sum ^9 _{m = 1} \sum ^9 _{n = 1}
k_{lmn} x_l x_m x_n \; ,
\label{basic eq. 7}
\end{equation}
where $n_{\rm H}$ denotes the total number density of hydrogen nuclei
and $ x_i \, \equiv n_i / n_{\rm H} $ is the relative number density 
of the $i$-th species.
The reaction rate coefficients,  $k_{jk}$ and $k_{lmn} $, 
are given in Table 1.


\begin{deluxetable}{llll} 
\tablecolumns{4}
\tablecaption{REACTION RATE COEFFICIENTS 
\label{table:physical_quantity_b1}}
\tablehead{
&\colhead{Reactions} & \colhead{Rate Coefficients} & Reference }
\startdata
  (1) & $ {\rm H}  + e \rightarrow {\rm H^+}  +  2 e  $
& $ k_1  = 5.85 \times 10^{-11}  T^{0.5} \exp (-157809.1/T)
(1 + T _5 {}^{0.7}) ^{-1}$ & 1   \nl
  (2) & ${\rm He}  +  e \rightarrow {\rm He ^+} + 2 e$
&$ k_2  = 2.38 \times 10^{-11} T ^{-0.5} \exp (-285335.4 /T)
(1 +  T_5 {}^{0.5})^{-1} $ & 1 \nl
  (3) & ${\rm He^+} + e  \rightarrow  {\rm He^{++}}  + 2 e $
& $ k_3  =  5.68 \times 10^{-12} T^{0.5} \exp (-631515.0 /T) 
(1 +  T_5 {}^{0.5})^{-1} $ & 1 \nl
  (4) &$ {\rm H^+}  +  e \rightarrow {\rm H}  +  h \nu  $
& $ k_4 = 8.40 \times 10^{-11}  T^{-1/2} T_3 {}^{-0.2} 
(1 + T_6 {}^{0.7}) ^{-1}$ & 1   \nl
  (5) & $ {\rm He ^+} + e \rightarrow {\rm He} + h \nu $
& $  k_5  =  1.9 \times 10^{-3} T ^{-1.5} \exp (-470000/T) 
(1  +  \exp (-94000/T)  $ & 1 \nl 
& & $\hspace{5mm} +  1.50 \times 10^{-10} T^{-0.6353} $  & \nl
  (6) & ${\rm He ^{++}}  +  e  \rightarrow  {\rm He ^+} \, + 
\, h \nu $& $ k_6  = 3.36 \times 10^{-10} T^{-0.5} T_3 {}^{-0.2} 
(1 + T_6 {}^{0.7}) ^{-1}$ & 1   \nl
  (7) & $ {\rm H}  + {\rm H} \rightarrow {\rm H} + 
{\rm H^+}   +  e $ & $ k_7  =  1.7 \times 10^{-4} k_1 $ & 2 \nl
  (8) &${\rm H} +  e  \rightarrow {\rm H ^-}  + h \nu $
&  $k_8  =  \left\{\begin{array}{ll}
1.0 \times 10^{-18} T & T  \le  1.5 \times 10^4 {\rm K} \\
{\rm dex} \left[-14.10 + 0.1175 \log T \right. & \\
\hspace{0.5cm}\left. -  9.813 \times 10^{-3} \right] (\log T)^2] & 
T \, > \, 1.5 \times 10^4 {\rm K} \end{array} \right. $
 & 3 \nl
  (9) &$ {\rm H}  +  {\rm H ^-}  \rightarrow {\rm H _2}  + e $ 
& $ k_9  =  \left\{\begin{array}{ll}
1.3 \times 10^{-9}  & T  \le 10^4 {\rm K}  \\
{\rm dex}\left[-8.78  +  0.113 \log T  \right. & \\
\hspace{0.5cm} \left. -  3.475 \times 10^{-2} (\log T)^2 \right]  & 
T  >  10^4 {\rm K}  \end{array}\right.
$ & 3 \nl
  (10) & $ {\rm H }  + {\rm H^+} \rightarrow  {\rm H_2 ^+} 
\, + h \nu $ & $k _{10} = \left\{\begin{array}{ll}
1.85 \times 10^{-23} T^{1.8} & T  \le  6.7 \times 10^3 {\rm K} \\
5.81 \times 10^{-16} & \\
\hspace{0.2cm} \times (T/56200)^{[-0.6657 \log (T/56200)]} & 
 T  >  6.7 \times 10^3 {\rm K} \end{array}\right. $ & 3 \nl
  (11) & ${\rm H_2 ^+} +  {\rm H} \rightarrow {\rm H_2} + 
{\rm H^+}$ & $ k_{11} =  6.4 \times 10^{-10} $ & 4 \nl
  (12) & ${\rm H_2}  + {\rm H} \rightarrow 3 {\rm H}$ &
$ k_{12} \; \mbox{[see equation (5) in reference]} $ & 5 \nl
  (13) & ${\rm H_2} + {\rm H^+}  \rightarrow  {\rm H_2 ^+} 
\, + \, {\rm H} $ & $k_{13} \, = \, 2.4 \times 10^{-9} 
\exp (-21200/T) $ & 3 \nl
  (14) & ${\rm H_2} + e  \rightarrow  {\rm H} + {\rm H^-} $
& $k_{14} = 2.7 \times 10^{-8} T^{-1.5} \exp (-43000/T) $ & 6 \nl
  (15) & ${\rm H_2} + e \rightarrow  2 {\rm H} + e $
& $k_{15} = 4.38 \times 10^{-10} T ^{0.35} \exp
(-102000/T) $ & 3\nl 
  (16) & ${\rm H_2} \, + \, {\rm H_2} \, \rightarrow \, 2 {\rm H}
\, + \,  {\rm H_2} $ & $k_{16} 
\; \mbox{[see equation (5) in reference]} $ & 5 \nl
  (17) & ${\rm H^-} \, + \, e \, \rightarrow \, {\rm H} \, + \, 2 e $
& $ k_{17} \, = \, 4.0 \times 10^{-12} \exp (-43000/T) $ & 3 \nl
  (18) & ${\rm H^-} \, + \, {\rm H} \, \rightarrow \, 2 {\rm H} \, + \, 
e $ & $k_{18} \, = \, 5.3 \times 10^{-20} T^{2.17} \exp
(-8750/T) $ & 3 \nl
  (19) & ${\rm H^-} \, + \, {\rm H^+} \, \rightarrow \, 2 {\rm H} $ 
& $ k_{19} \, = \, 7.0 \times 10^{-7} T^{-0.5} $ & 7 \nl
  (20) & ${\rm H^-} \, + \, {\rm H^+} \, \rightarrow \, {\rm H_2 ^+}
\, + \, e $ & $k_{20} \, = \, \left\{\begin{array}{ll}
10^{-8} T^{-0.4} & T \, \le \, 10^4 {\rm K} \\
4 \times 10^{-4} T^{-1.4} \exp (-15100/T) & T \, > \, 10^4 {\rm K} 
\end{array} \right. $ & 3 \nl
  (21) & ${\rm H_2 ^+} \, + \, e \, \rightarrow \, 2 {\rm H} $
& $ k_{21} \, = \, 1.68 \times 10^{-8} (T/300) ^{-0.29} $ & 8 \nl
  (22) & $ {\rm H_2 ^+} \, + \, {\rm H^-} \, \rightarrow \, {\rm H}
\, + \, {\rm H_2} $ & $ k_{22} \, = \, 5.0 \times 10^{-6} T^{-0.5}$ &7 \nl
  (23) & $ 3 {\rm H} \, \rightarrow \, {\rm H_2} \, + \, {\rm H} $ 
& $ k_{23} \, = \, 5.5 \times 10^{-29} T^{-1} $ & 9 \nl
  (24) & $2 {\rm H} \, + \, {\rm H_2} \, \rightarrow \, 2 {\rm H_2} $
& $ k_{24} \, = \, k_{23} / 8 $ & 9 \nl
\enddata
\tablenotetext{a}{The units of rate coefficients are taken to be
cm$^3$ s$^{-1}$ for two-body reactions and cm$^6$ s$^{-1}$ for 
three-body reactions}
\tablecomments{(1) Cen 1992; (2) Palla et al. 1983; 
(3) Shapiro \& Kang  1987; (4) Karpas, Anicich, \& Huntress 1979;
(5) Lepp \& Shull 1983; (6) Hirasawa 1969; (7) Dalgarno \& Lepp 1987;
(8) Nakashima, Takayi \& Nakamura 1987; 
(9) Palla et al. 1983}
\end{deluxetable}

In the models calculated in this paper,
the gas temperature does not become far above 10$^4$ K.
Thus the cooling is dominated by
contributions from H at $T \sim 10^4 $ K and H$_2$
at $ T <  10^4 $ K.
We therefore adopt the cooling rate in the form of 
$ \Lambda _{\rm net} = \Lambda _{\rm H} + 
 \Lambda _{\rm H_2} +  \Lambda _{\rm chem} $, 
where $\Lambda _{\rm H} $,  $\Lambda _{\rm H_2}$, and
$\Lambda _{\rm chem}$ denote contributions from H, H$_2$, and
chemical reactions, respectively.
$\Lambda _{\rm H}$ includes the cooling of H atoms due to 
recombination, collisional ionization, and collisional excitation
(Cen 1992; see also Black 1981).
$\Lambda _{\rm H_2}$ includes the cooling of H$_2$ due to 
rotational and/or vibrational excitations
(Lepp \& Shull 1983; Haiman et al. 1996), 
collisional dissociation (Lepp \& Shull 1983), 
and heating due to H$_2$ formation (Shapiro \& Kang 1987).
$\Lambda _{\rm chem}$ includes  the cooling due to 
chemical reactions in Table 1 (Shapiro \& Kang 1987).
For each contribution, we use the analytic formula expressed 
in each reference.
[It is claimed that the cooling rate by Lepp \& Shull (1983) 
is overestimated in relatively low densities ($n_H<1 {\rm cm}^{-3}$),
compared to the rate by Hollenbach \& McKee (1989).
But, they are in a good agreement with each other in higher densities
relevant to the present issue (Galli \& Palla 1998).]

When the H$_2$ density exceeds $\sim 10^{10} $ cm$^{-3}$,
the H$_2$ line emission becomes optically thick (Palla et al. 1983).
To take account of this effect, we modify the H$_2$ line cooling 
with using the photon escape probability (Castor 1970; 
Goldreich \& Kwan 1974): 
$\Lambda _{\rm H_2 line} = \beta \Lambda _{\rm H_2 line, thin}$ 
and $\beta = (1  - e ^{-\tau _R})/\tau _R$,
where $\Lambda _{\rm H_2 line, thin}$ is the cooling function in
optically thin regime, $\beta$ is the photon escape probability, 
and $\tau _R$ is the Rossland mean opacity for the H$_2$ line emission.
To evaluate $\tau _R$, we determined the level populations
with the method of Palla et al. (1983).
This type of $\beta$ is applicable rigorously 
when the gas flows supersonically and its velocity monotonically changes 
in proportion to the radius (Castor 1970; Goldreich \& Kwan 1974).
However, the evolution has turned out to depend weakly on the form of $\beta$.
Using other types of $\beta$, e.g., $ \beta \, = \,  e ^{-\tau _R}$, 
we have recalculated the evolution and reached basically the same results.

\subsection{A Model for a Filamentary Primordial Cloud}
\label{subsec:formation}

In a gravitational instability scenario for the formation 
of the first structures,
a cosmological density perturbation larger than the Jeans scale 
at the recombination epoch forms a flat pancake-like disk.
This process has been extensively 
studied by many authors (e.g., Zel'dovich 1970; Sunyaev \& Zel'dovich 1972;
Cen \& Ostriker 1992a, 1992b; Umemura 1993).
Although the pancake formation is originally studied by Zel'dovich (1970)
in the context of the adiabatic fluctuations in baryon or
hot dark matter-dominated universes, 
recent numerical simulations have shown that 
such pancake structures also emerge in the CDM cosmology
(e.g., Cen et al. 1994).
Thus, the pancakes are thought to be a ubiquitous feature
in gravitational instability scenarios.

In the bottom-up theory like a CDM model, 
the first collapsed objects at $z \sim 10 - 100 $
are expected to have the masses of $\sim 10^{5-7} M _\odot$.
In these clouds the gas is heated above $ T > 10^{3 - 4} $ K by shock.
Therefore, just after the shock formation, the cooling timescale 
is likely to be shorter than the free-fall one; 
the temperature of the pancake reduces to the value 
at which the cooling timescale is comparable to the free-fall one
(e.g., Haiman et al. 1996; see also Yoneyama 1972).
The temperature of the pancake is then estimated to be 
$ T \simeq 200 - 1000 $ K for $y_{\rm H_2} = 10^{-4} \sim 10^{-3}$.
Thereafter, the pancake is likely to fragment into filamentary clouds
which have nearly the same temperature as that of the parent pancake
(Miyama et al. 1987a, 1987b; Uehara et al. 1996).
In the following we describe a model of a filamentary cloud formed 
by fragmentation of the pancake. 

As a model of a filamentary gas cloud, we assume an infinitely 
long cylindrical gas cloud for simplicity.
At the initial state, we assume the gas to be quiescent and isothermal.
We also assume that 
H$^-$, H$_2 ^+$, He$^+$, and He$^{++}$ 
do not exist at the start
and the relative abundances of other species are spatially uniform 
for simplicity.
The density distribution in the radial direction is set to be
\begin{equation}
\rho \, = \, \rho _0 \left( 1 \, + \, 
{r^2 \over f R_{\rm fil} ^2} \right)^{-2} \; ,
\end{equation}
and
\begin{equation}
R_{\rm fil} \, = \, \sqrt{2 k T_0 \over \pi G \rho _0 \mu} \; ,
\end{equation}
where $\rho_0$ and $T_0$ are respectively the central density and
the initial gas temperature, and $f$ represents the degree of deviation 
from the equilibrium state.
A model is specified by the following five parameters: $\rho_0$,
$T_0$, $f$, the electron number fraction $x_e$, and 
the H$_2$ number fraction $x_{\rm H_2}$.
When $f \, = \, 1$, the cloud is just in hydrostatic equilibrium; 
the density distribution accords with that of an isothermal 
filamentary gas cloud in equilibrium 
(Stod\'o\l kiewicz 1963).  
In this paper, we restrict the parameter $f$
to $ f \, \ge \, 1 $ since we are interested in the evolution of 
the collapsing clouds.

For the above model, 
the mass per unit length (line mass) is given by
\begin{eqnarray}
l_0 & = & \int ^\infty _0 2 \pi \rho r dr \, = \, \pi R_{\rm 
fil} ^2 \rho _c f  \nonumber \\
& = & {2 k T_0 \over \mu G} f 
= 2.2 \times 10^3 \left({f \over 2}\right) M_\odot {\rm pc^{-1}}
\left({T_0 \over 1000 {\rm K}}\right) \; .
\end{eqnarray}
Note that the line mass of the equilibrium filamentary cloud
depends only on the temperature.
When the filamentary cloud forms through gravitational fragmentation of
the pancake, its line mass is estimated to be 
$ l \sim 2 \rho _d \lambda_m H_d \sim 4 k T/(\mu G)$, which 
is twice that of the equilibrium cloud; i.e., the typical value of $f$ 
is evaluated to be $f = 2$ (Miyama et al. 1987a).
Here $\rho_d$, $\lambda_m$, and $H_d$ are the mass density of the
parent pancake, the wavelength of the most unstable linear
perturbation, and the half thickness of the pancake, respectively.
We thus adopt $f =2$ for a typical model.

\subsection{Model Parameters and Numerical Method}

As shown in \S \ref{subsec:formation}, the initial state
is specified by the parameters $n_0 (\equiv \rho_0/\mu$), $T_0$, $f$,
$x_e$, and $x_{\rm H_2}$.
We examine 60 models, by choosing the parameters $n_0$, $T_0$, and $f$ 
to be $ n_0 \, = \, 10^2 $, $10^3$, $\cdots$, $10^6$ cm$^{-3}$, 
$ T_0 \, = \, 10^2 $, $5 \times 10^2$, $10^3$, $10^4$ K,
and $ f \, =\, 2 $, 4, 10, respectively.
For the parameters $x_e$ and $x_{\rm H_2}$, 
we adopt $ x _e   = 5 \times 10^{-5} $ and 
$x_{\rm H_2} = 10^{-4} $ for the models with $ T_0 \, < \, 10 ^4 $ K.
The value of $x_e = 5 \times 10^{-5} $ 
is adopted from the calculation of
Peebles (1968) for the residual post-recombination ionization.
[We have found that the results do not depend upon 
$x_e$ as far as $x_e \gtrsim 10^{-7}$, because
the free electrons could quickly recombine 
to a level of $x_e < 10^{-7}$ in the course of the collapse.
(see also Haiman, Rees, \& Loeb 1996, 1997).]
The other abundances are determined by 
the conservation of mass and charge.
For the models with $T_0 = 10^4 {\rm K} $, 
we determine the abundances from the statistical equilibrium
with e, H, H$^+$, He, He$^+$, and He$^{++}$.

The hydrodynamic equations (\ref{basic eq. 1}) 
- (\ref{basic eq. 6}) are solved numerically 
by using the second-order upwind scheme based on 
Nobuta \& Hanawa's (1998) method.  
This scheme is an extension of Roe's (1981) method 
to the gas having non-constant $\gamma$. 
[See Nobuta \& Hanawa (1998) for more details and the test of the code.]
The rate equations (\ref{basic eq. 7}) are 
solved numerically with the LSODAR (Livemore Solver for
Ordinary Differential equations with Automatic method switching for
stiff and non-stiff problems) coded by L. Petzold and A. Hindmarsh.

As for the boundary condition, we take the fixed boundary at $r \, = \, 
R_{\rm max} $, where $R_{\rm max}$ is the maximum radial coordinate
in the computational domain. In all the models we have taken
$ R_{\rm max}$ to be about ten times greater than the effective radius
 of the filamentary cloud, $f^{1/2} R_{\rm fil}$. 
The effect of the fixed boundary is very small.
This is because the density is much lower near the outer boundary 
than at the center, i.e.,
$\rho \, \lesssim \, 10^{-4} \rho _c $ for all runs.
(In fact, we have calculated the case with larger $R_{\rm max}$
and have confirmed that the numerical results are not changed.)

The numerical grids are non-uniformly distributed
so as to enhance the spatial resolution near the center.
The grid spacing increases by 5 \% for each grid zone
with increasing distance from the center.
As shown in \S \ref{sec:results}, the characteristic scale shortens 
in the central high density region as the collapse proceeds.
The spatial resolution thus becomes poor near the center.
To compute the further contraction with the sufficient spatial resolution,
we pursue the subsequent evolution with refining grids.
Whenever the number of grid points becomes less than $\sim 20$
within the radius of the half-maximum density ($\rho = 0.5 \rho _c$),
we increase the number of grid points 
and then reposition them on the refined grids 
in the whole computation region ($ 0 \leq r \leq R_{\rm max}$).
The physical variables at the new grid points are determined
by linear interpolation. This technique allows us to 
pursue the dynamical evolution over more than the tenth order of
magnitude in the central density.

\section{Numerical Results}
\label{sec:results}

\subsection{Clouds with Intermediate and High Initial Temperatures
($ T_0 \gtrsim 500$ K)}
\label{subsec:500K}

We have examined the evolution of various models with relatively 
high initial temperature ($ T_0 \gtrsim 500$ K).
As a result, the way of the evolution has turned out to be 
almost insensitive to the model parameters.

\begin{figure}
\plotone{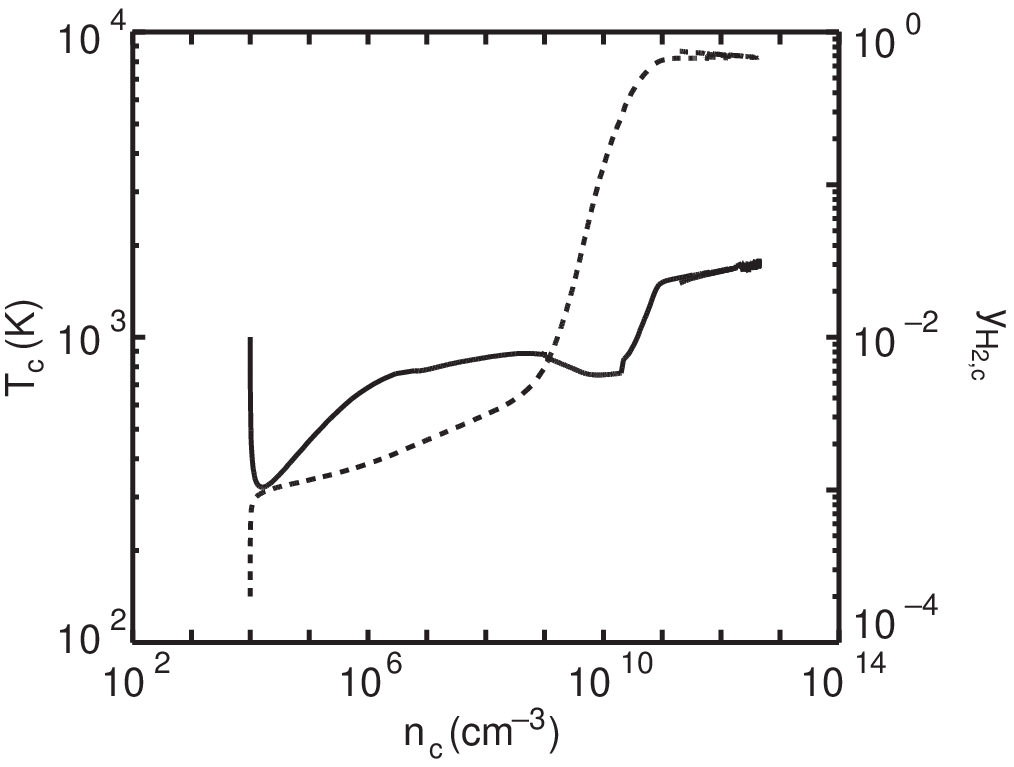}
\caption{Evolution of the temperature ($T_c$, {\it solid curves}) 
and the H$_2$ abundance
($y_{\rm H_2, c}$, {\it dashed curves}) at the center for the model with 
$(n_0, T_0, f)=(10^4 \, {\rm cm}^{-3},\ 10^3 \, {\rm K}, \ 2)$. 
The first prompt decrease of the temperature is due to
the H$_2$ cooling.
Afterwards, until the central density reaches $\sim$ 10$^{11}$ cm$^{-3}$, 
the temperature keeps a nearly constant value of $\sim $ 800 K
over the fifth order of magnitude in density.
The H$_2$ abundance steeply rises around
$n_c \sim 10^{9} $ cm$^{-3}$ and reaches $y_{\rm H_2} \sim 1 $
at the stage at which $n_c \sim 10^{11} $ cm$^{-3}$.
At $n_c \sim 10^{11} $ cm$^{-3}$, the cloud becomes optically thick
against the H$_2$ lines, and consequently the temperature rises to $\sim$ 1500 
K.
Thereafter the cloud contraction decelerates near the center
and the cloud weakly oscillates around 
its quasi-static equilibrium state.
\label{fig:rho-t}}
\end{figure}

As a typical example,
we first show the evolution of the model
with $(n_0, \, T_0, \, f) $ = $(10^4 \, {\rm cm}^{-3}, \, 
10^3 \, {\rm K}, \, 2) $.
Figure \ref{fig:rho-t} shows the evolution of 
the temperature ($T_c$) and the mass fraction of H$_2$ at the center 
($ y_{\rm H_2} \equiv 2 x_{\rm H_2}$) 
as a function of the central density $n_c$.
During the early evolution, H$_2$ forms mainly through 
the H$^-$ process [eqs. (\ref{eq:hm1}) and (\ref{eq:hm2})].
Because of the effective cooling
by the rotational transitions,
the temperature first descends promptly to $ T \sim 300 $ K. 
When the central density increases to $n_c \, \sim \, 10^{6} $ cm$^{-3}$, 
the cloud collapses nearly isothermally, 
staying the temperature at $ T \sim 800 $ K
until the central density reaches $n_c \, \sim \, 10^{10-11} $ cm$^{-3}$.
The temperature evolution can be then approximated 
by $ T _c \, \propto \, n_c ^{0.15} $ 
as far as  $ n_c \, < \, 10^8 $ cm$^{-3}$.
When the density exceeds $ \sim 10^8$ cm$^{-3}$, 
the hydrogen molecules form acceleratively
through the three-body reactions 
[eqs. (\ref{eq:3h process1}) and (\ref{eq:3h process2})].
After the central density increases to $n_c \, \sim \, 10^{11} $ cm$^{-3}$,
the hydrogen gas is almost completely processed into molecules.
Then the cloud becomes optically thick 
against the H$_2$ lines and the temperature rises up to $\sim$ 1500 K.
At this stage, the cooling time is $50-100$ times longer than 
the free-fall time.
Thereafter the collapse decelerates  near the center and
the cloud weakly oscillates around its quasi-static equilibrium state.

In Figures \ref{fig:density etc}a - \ref{fig:density etc}d,
the time variations of the distributions of
the density, the temperature, the H$_2$ mass fraction, 
and the contraction time are plotted at several dynamical stages.
The contraction time is a dimensionless one, which is defined as 
$ t _{\rm cont} (r)  =  r /\left[v_r (r) t_{\rm ff} \right] $,
where $t_{\rm ff}$ is the free-fall time at the center, 
$t _{\rm ff} \equiv (4 \pi G \rho _c ) ^{-1/2} $.
Since the gravitational force is perceptibly greater than 
the pressure force at the initial state, the cloud collapses 
nearly in a free-fall time.
When the central density reaches 
$n _c  \sim 10^8 $ cm$^{-3}$ ($t=1.434 \times 10^6$yr), 
the shock arises around $ r  \sim 3 \times 10^{-2}$ pc, 
which is characterized by a jump in density and temperature.
The shock surface separates the cloud into two parts
which correspond to mutually different dynamical states, that is,  
a denser spindle and a diffuse envelope.  
During the contraction, the temperature in the spindle 
keeps nearly constant around $800 $ K 
until the cloud becomes optically thick against the H$_2$ lines.
In the isothermal contraction phase, the outer parts in the
spindle exhibit a power-law density distribution, $\rho \propto r^{-2}$.
It indicates that the spindle contracts in a self-similar manner. 
Actually, we have found a similarity solution, which is presented
in the Appendix, and it has turned out that
the newly found similarity solution well reproduces the numerical results. 
(See the Appendix for the further detail)
At the shock front, the temperature rises up to $T \sim 1500 $ K.
The contraction time in the spindle is much shorter than 
that in the envelope.
Thus the spindle collapses almost independently of the envelope.
When the cloud becomes optically thick against the H$_2$ lines
at the center, the pressure force overwhelms the gravity, so that 
the second shock forms around $ r \sim 2 \times 10^{-4}$ pc.
After this stage the contraction time of the spindle becomes
much longer than $t_{\rm ff}$ and the collapse promptly
decelerates.

Figure \ref{fig:line mass} shows the evolution of 
the line mass of the spindle
as a function of the central density.
We define the line mass as the density integrated
over $r$ up to the radius at which the density goes down to 
one tenth of the central density, 
\begin{equation}
 l _{\rm sp} \, = \, \int _0 ^{r (\rho = 0.1 \rho_c)} 2 \pi r \rho dr \; .
\end{equation}
It is worth noting that during the contraction, the line mass of the spindle
keeps a nearly constant value, which is $\sim 1 \times 10^3 M_\odot$ pc$^{-1}$.
This value is close to the line mass of 
an isothermal filamentary cloud in equilibrium for the temperature
of 800 K, i.e., 
$l_0 = {2 k T/ \mu G} = 0.9 \times 10^3 M_\odot {\rm pc^{-1}}
\left({T / 800 {\rm K}}\right)$.
It implies that the gravitational force 
nearly balances with the pressure force in the spindle.

We have found that the evolution is almost independent 
of the initial parameters, $n_0$, $T_0$, and $f$, 
for the models with $T_0 \gtrsim 500$ K.

\subsection{Clouds with Low Initial Temperature
($ T_0 \sim 100$ K)}

In this subsection, we examine the evolution of models
with relatively low initial temperature; $T_0 \, = \, 100 $ K.
We find that the evolution in the case of low initial temperature 
depends mainly on the value of $f$.
In Figures \ref{fig:low temp}a and  \ref{fig:low temp}b,
two models with 
$(n_0, T_0, f) $ = $(10^4 \, {\rm cm}^{-3}, \, 10^2 \, {\rm K}, \, 2) $ 
and $(10^4 \, {\rm cm}^{-3}, \, 10^2 \, {\rm K}, \, 10) $ are compared.
For both models the early evolution is similar;
the clouds collapse adiabatically,
$T \propto \rho ^{2/3} $, because the H$_2$ cooling 
is not effective for $ T \lesssim 300 $ K.
However the later evolution is quite different.
For the model with a small $f$, the collapse almost ceases 
until the central density reaches $10^5$ cm$^{-3}$.
In this case, the contraction time, which is nearly equal to the cooling time, 
is a hundred times as long as the free-fall time.
Hence, the cloud becomes close to the hydrostatic equilibrium
with no efficiency of cooling.
On the other hand, for the model with a large $f$,
the evolution is similar to that of the model 
with $ T_0 \gtrsim 500 $ K. When the cloud temperature exceeds  500 K,
the collapse proceeds nearly isothermally due to the H$_2$ line cooling.
In all the models with $ T_0 = 100$ K,
it has turned out that the evolution proceeds in a similar way 
dependent upon the value of $f$.

\section{Fragmentation of Primordial Gas Clouds}
\label{sec:fragmentation}

As shown in the previous section, if $T _0 \gtrsim 500 $ K or
$ f \gtrsim 10$, the filaments collapse quasi-statically, whereas 
the filaments would not continue to contract if $T _0 \lesssim 100 $ K 
and $ f \lesssim 2 $. In the latter case, the filament would be
weakly unstable because the length of the realistic filament
is shorter than the wavelength of the most unstable mode.
Consequently, the filament is likely to contract along the axis without
fragmentation. Also, as mentioned in \S \ref{subsec:formation}, 
the temperature of the initial filaments is 
estimated to be $ T \sim 200 - 1000 $ K in a CDM model.
Hence, such low temperature clouds are hardly expected in a CDM scenario.
Thus, here we focus on the former case, i.e., a collapsing 
filament, and consider the hierarchical fragmentation.

The density of a collapsing filament is enhanced by more than eighth order of
magnitude, with the constant temperature of $ T\sim 800 $ K, until
the cloud becomes opaque. Thus, the filament could fragment 
into denser lumps.
To diagnose the fragmentation process, we apply a linear theory 
to the numerical results
and thereby estimate the length and mass scales of fragments.

According to the linear theory of an isothermal hydrostatic filament,
the perturbation grows most rapidly when its wavelength is about four
times longer than the effective cloud diameter.
The wavelength and growth rate of the most unstable perturbation
depend on the density and temperature of the cloud.
Therefore, they change according as the collapse proceeds.
In our model of \S \ref{subsec:500K}, the contraction timescale 
($t_{\rm cont}$) is longer by a factor of more than five 
than the free-fall time at the center ($t_{\rm ff}$) 
after the central density reaches $\sim 10^6$ cm$^{-3}$.
According to Inutsuka \& Miyama (1992), 
when $t_{\rm cont} \gtrsim 10 t_{\rm ff}$, the dispersion relation for 
the collapsing filament is approximated by that of the hydrostatic
filament whose scale height is the same as 
the temporal one of the collapsing cloud, 
and thus the growth of the density perturbation is
approximated by 
\begin{equation}
\delta \rho (t) / \rho (t) \, = \, A \exp \left(i k _z z \, - \, 
\int _0 ^t i \omega (k_z, \, t') dt' \right),
\end{equation}
during its linear growth phase, 
where $k_z$ and $\omega$ denote the wavenumber and the frequency 
of the perturbation, respectively.
We thus apply the above equation to our model of \S \ref{subsec:500K}.
For the growth rate ($- i \omega$), we use a fitting formula
obtained by Nakamura, Hanawa, \& Nakano (1993).
$A [\, = \, A (k_z)]$ denotes the initial amplitude
of the perturbation. We take $A$ in the form of the power law
$A = A_0 (k_z/ k_{z, 0})^p $.
When the amplitude of the unstable density perturbation
becomes greater than unity ($\delta \rho / \rho \gtrsim 1 $),
the evolution of the perturbation becomes nonlinear and
the cloud will breaks into pieces.
It is expected that the evolution of the perturbation depends on 
the index $p$ and the amplitude $A_0$.

Assuming $p=1$ or $p=-1$, we have calculated the time evolution of 
the density contrast in the filamentary cloud.
The results are shown in Figure \ref{fig:growth}. 
The solid curves show the density contrast  
at  (1) $t \, = \, 0$ yr ($n _c \, \simeq \, 10^4 $ cm$^{-3}$), 
(2) $1.384 \times 10^5$ yr (10$^{7}$ cm$^{-3}$), 
(3) $1.477 \times 10^5$ yr (10$^{9}$ cm$^{-3}$), 
(4) $1.500 \times 10^5$ yr (10$^{11}$ cm$^{-3}$), 
and (5) $1.503 \times 10^5$ yr ($3 \times 10^{12}$ cm$^{-3}$),
respectively.
Here, we have taken $(A_0, k_{z,0}) = (1.0 \times 10^{-3},  
1.26 \times 10^2 {\rm pc^{-1}})$.
When $n \lesssim 10^{10 - 11}$ cm$^{-3}$, the wavelength of 
the fastest growing perturbation is about a hundred times as long as 
the effective cloud diameter, because of the cumulated growth rate.
Hence, the cloud is likely to tear into long filaments
before the central density reaches $n_c \sim 10^{11}$ cm$^{-3}$, 
if the fluctuations have enough initial amplitudes to enter nonlinear stages.
The long filaments shrink further to form thinner filaments.
As the shrink proceeds, 
the wavelength of the fastest growing perturbation shortens.
Thus, the filaments may again tear into shorter and 
denser filaments. Consequently, hierarchical filamentary structures 
would form, if the initial amplitudes are larger than
0.001 as expected in a CDM cosmology.

When the central density exceeds $n_c \sim 10^{11}$ cm$^{-3}$,
the cloud becomes opaque against the H$_2$ lines
and the collapse quickly decelerates.
Hence the hierarchical fragmentation would be terminated.
Then the wavelength of the fastest growing perturbation 
becomes $ \lambda _* \, \sim \, 2 \times 10^{-3} {\rm pc}$,
which is comparable to that expected from 
the linear theory of an isothermal filament in equilibrium,
as shown by the growth rate in the ending stage in Figure 5.
Then, the cloud is most unstable 
against the perturbation whose wavelength is about four times
as large as the effective cloud diameter.
This wavelength is nearly independent of the power index $p$
as shown in Figure \ref{fig:growth}.
As a result, the filament is likely to finally fragment 
into dense cores, the typical mass of which is
$\lambda _* l_{\rm sp} \sim (2 \times 10^{-3} {\rm pc})
\times (1.5 \times 10^3 M_\odot {\rm pc^{-1}}) \sim 3 M_\odot$.
This is the lowest mass of fragments, which provides 
the minimum core mass for the first-generation stars.
The mass possibly increases by accreting the ambient gas.
If all the ambient gas within one wavelength of the 
fastest growing perturbation accretes onto the core, 
then the core mass increases to
$\lambda _* l_0 \sim 16 (f/2 )(T_0/1000 \ {\rm K}) M_\odot$.
Thus, the mass of the first stars is expected to be in the range of 
$3  M_\odot \lesssim M \lesssim 16 (f/2 )(T_0/1000 \ {\rm K}) M_\odot $.
It should be noted that the lower mass as well as the upper mass
is not sensitive to the amplitude of the power-law spectrum, because
the lower mass is basically determined by the micro process of the H$_2$
cooling and the upper mass depends only upon the initial temperature.

\section{Conclusions}
\label{sec:sec6}

In a wide range of the parameter space, 
we have numerically explored the thermal and dynamical evolution of 
filamentary primordial gas clouds with including nonequilibrium 
process for the H$_2$ formation.
We have found that for the models with $T_0 \gtrsim 500 $ K or 
$ f \gtrsim 10 $,
the collapse proceeds nearly isothermally due to the H$_2$ line 
cooling, staying the temperature at $T \sim 800 $ K.
During the contraction, the cloud structure is divided into 
two parts, i.e., a spindle and an envelope.
The line mass of the spindle keeps a nearly constant value
of $\sim 1 \times 10^3 M_\odot $  pc$^{-1}$
during the contraction from $10^{6}$ cm$^{-3}$ to $10^{11}$ cm$^{-3}$.
The outer part in the spindle exhibits a power-law density 
distribution as $\rho \propto r^{-2}$.
This behavior is well reproduced by a newly found self similar solution.
When the central density reaches $ n_c \gtrsim 10^{11} $ cm$^{-3}$, 
the cloud becomes opaque against the H$_2$ line cooling and 
reaches a quasi-static equilibrium state with keeping the line mass.
On the other hand, for the models with $T _0 \sim 100 $ K and 
$ f \lesssim 2 $, the cloud contraction is much slower
because the H$_2$ cooling is not effective.
The contraction immediately ceases.
In a CDM scenario, the former case (collapsing filaments)
would be more probable.

Applying a linear theory for the gravitational instability of 
collapsing filaments, we find that
the spindle is unstable against fragmentation 
during the quasi-static contraction.
The wavelength of the fastest growing perturbation 
($\lambda _{\rm m}$) lessens as the collapse proceeds.
Thus successive fragmentation could occur. 
When the central density exceeds $n _c \, \sim \, 10^{11}$ cm$^{-3}$,
the successive fragmentation may be suppressed 
since the cloud becomes opaque against the H$_2$ lines
and the collapse promptly decelerates.
Resultingly, the minimum value of $\lambda _{\rm m}$
is estimated to be $\sim 2 \times 10^{-3}$ pc.
The typical mass of the first stars is then expected to be 
$\sim 3 M_\odot$, which may grow up to $16 M_\odot $ by accreting
the diffuse envelope.
The present results may be relevant to the early evolution 
of primordial galaxies and the metal enrichment of the intergalactic space.

\acknowledgments

We are grateful to Drs. 
T. Nakamoto, and H. Susa for stimulating discussion.
We also thank an anonymous referee for valuable comments 
which improved the paper.
This work was carried out at the Center for Computational Physics
of University of Tsukuba.
This work was supported in part by the Grants-in Aid of the
Ministry of Education, Science, and Culture 
(09874055, 09740171, 10147205).

\newpage
\appendix

\section{Similarity Solutions of A Polytropic Filamentary Cloud}
\label{sec:app2}

As mentioned in \S 3, 
the spindle seems to collapse self-similarly
during the nearly isothermal contraction phase
from $ n_c \sim 10^6$ cm$^{-3}$ to 10$^{11}$ cm$^{-3}$; e.g.,
the density distribution of the spindle is characterized by
the power-law distribution of  $\rho \propto r^{-2}$. 
But this behavior is different from that of the similarity solution
($ \rho \propto r^{-4}$) of an isothermal filamentary cloud 
that is known so far
(Miyama et al. 1987a; see also eqs. [\ref{eq:iso1}] and [\ref{eq:iso2}]).
In this appendix we seek another type of similarity solutions 
in more generic equations of state.

We consider an infinitely long cylindrical cloud in which the density 
is uniform along the axis.
We assume the gas to be polytropic; $P  = K \rho ^\gamma$, where
$K$ is constant.
The equation of motion and the continuity equation are then described as 
\begin{equation}
{\partial v_r \over \partial t} + 
v_r {\partial v_r \over \partial r} + {2 G l \over r} +  
{1 \over \rho} {\partial P \over \partial r} =  0 \; ,
\label{ap2:eq1}
\end{equation}
\begin{equation}
{\partial l \over \partial t} + 2 \pi r \rho v_r = 0 \; ,
\label{ap2:eq2}
\end{equation}
and
\begin{equation}
{\partial l \over \partial r}  -  2 \pi r \rho  =  0\; ,
\label{ap2:eq3}
\end{equation}
where $l$ denotes the line mass (mass per unit length) 
 contained within a radius $r$.

We now look for a similarity solution of the form
\begin{equation}
x = {r \over a t} \; ,
\label{ap2:eq4}
\end{equation}
\begin{equation}
\rho (r,t) = {\tilde{\rho} (x) \over 2 \pi G t^2} \; ,
\label{ap2:eq5}
\end{equation}
\begin{equation}
v_r (r, t) = a \tilde{v}_r (x) \; ,
\label{ap2:eq6}
\end{equation}
\begin{equation}
l (r, t) = {a^2 \over 2 G} \tilde{l} (x) \; ,
\label{ap2:eq7}
\end{equation}
and
\begin{equation}
a  = t^{1-\gamma} \sqrt{K(2 \pi G)^{1-\gamma}}  \; ,
\label{ap2:eq8}
\end{equation}
where the physical variables with a tilde denote
those in the similarity coordinate, $x$.
Substituting equations (\ref{ap2:eq4}) $-$ (\ref{ap2:eq8}) into
equations (\ref{ap2:eq1}) $-$ (\ref{ap2:eq3}), we obtain
\begin{equation}
{d \tilde{\rho} \over dx}  = {Y(x, \tilde{\rho}, \tilde{v}_r)
\over X(x, \tilde{\rho}, \tilde{v}_r)} \; ,
\label{ap2:eq9}
\end{equation}
\begin{equation}
{d \tilde{v}_r \over dx} = {Z(x, \tilde{\rho}, \tilde{v}_r)
\over X(x, \tilde{\rho}, \tilde{v}_r)} \; ,
\label{ap2:eq10}
\end{equation}
\begin{equation}
X(x, \tilde{\rho}, \tilde{v}_r) = 
[\tilde{v}_r  - (2-\gamma) x]^2  - \gamma
{\tilde{\rho}}^{\gamma -1} \; ,
\label{ap2:eq11}
\end{equation}
\begin{equation}
Y(x, \tilde{\rho}, \tilde{v}_r) = - \left[\tilde{v}_r -
(2-\gamma) x \right]^2
\left[{\tilde{\rho} ^2 \over \gamma - 1}  + (2 x - \tilde{v}_r)
{\tilde{\rho} \over x}\right]  -  (\gamma - 1) \tilde{\rho} 
\tilde{v}_r  \; ,
\label{ap2:eq12}
\end{equation}
and
\begin{equation}
Z(x, \tilde{\rho}, \tilde{v}_r) = - [\tilde{v}_r - (2-\gamma) x]^2
{\tilde{\rho} \over \gamma - 1} + (\gamma -1) \tilde{v_r}
[\tilde{v}_r - (2-\gamma) x] + {\gamma
\tilde{\rho} ^{\gamma -1} \over x} (\tilde{v}_r - 2 x)  \; .
\label{ap2:eq13}
\end{equation}
Note that equations (\ref{ap2:eq9}) through (\ref{ap2:eq13}) 
have a singularity at $\gamma  = 1$ (A polytropic sphere has 
a singularity at $\gamma  = 4/3$. This is fundamentally the same
as the dynamical stability condition of the sphere.).

One of the solutions of equations (\ref{ap2:eq9}) 
through (\ref{ap2:eq13}) is expressed as 
\begin{equation}
\tilde{v}_r = 0 
\end{equation}
and 
\begin{equation}
\tilde{\rho} = \left[{(2  - \gamma)^2 \over 
2 \gamma (1 - \gamma)} \right] ^{1 \over \gamma - 2} 
x^{-{2 \over 2  -  \gamma}} \; .
\end{equation}
In this solution the gas is at rest and the density diverges at the
origin.  This solution corresponds to a singular equilibrium solution
for a polytropic filamentary cloud.

A regular solution at $ x  = 0$ has an asymptotic behavior,
\begin{equation}
{d \ln \tilde{\rho} \over d \ln x} = - {2 \over 2 - \gamma}
\end{equation}
and
\begin{equation}
{d \ln \tilde{v}_r \over d \ln x} 
= - {2 (\gamma -1) \over 2 - \gamma} \; .
\end{equation}
Thus, when the value of $\gamma$ is nearly equal to unity, 
the density and velocity distributions  follow 
$\rho \rightarrow  r^{-2}$ and 
$v_r  \rightarrow {\rm const.}$ 
in the outer region of $|x| \gg 1 $.
This behavior is quite similar to that of a sphere 
(Larson 1969; Penston 1969).
Note that this similarity solution exists only for $ \gamma < 1 $. 
This is essentially the same as the dynamical stability condition
of the polytropic cylinder.
This similarity solution has different characteristics from 
a singular case of $\gamma = 1$.
In the similarity solution of $\gamma = 1$, 
the density and velocity distributions 
are given as (see Miyama et al. 1987a)
\begin{equation}
\tilde{\rho}  =  {4 \over (1 +  x^2)^2} \; ,
\label{eq:iso1}
\end{equation}
and
\begin{equation}
\tilde{v}_r  =  x \; ,
\label{eq:iso2}
\end{equation}
which are proportional to $x^{-4}$ and $x$, respectively, in the
region of $|x| \gg 1$.

\begin{figure}
\plottwo{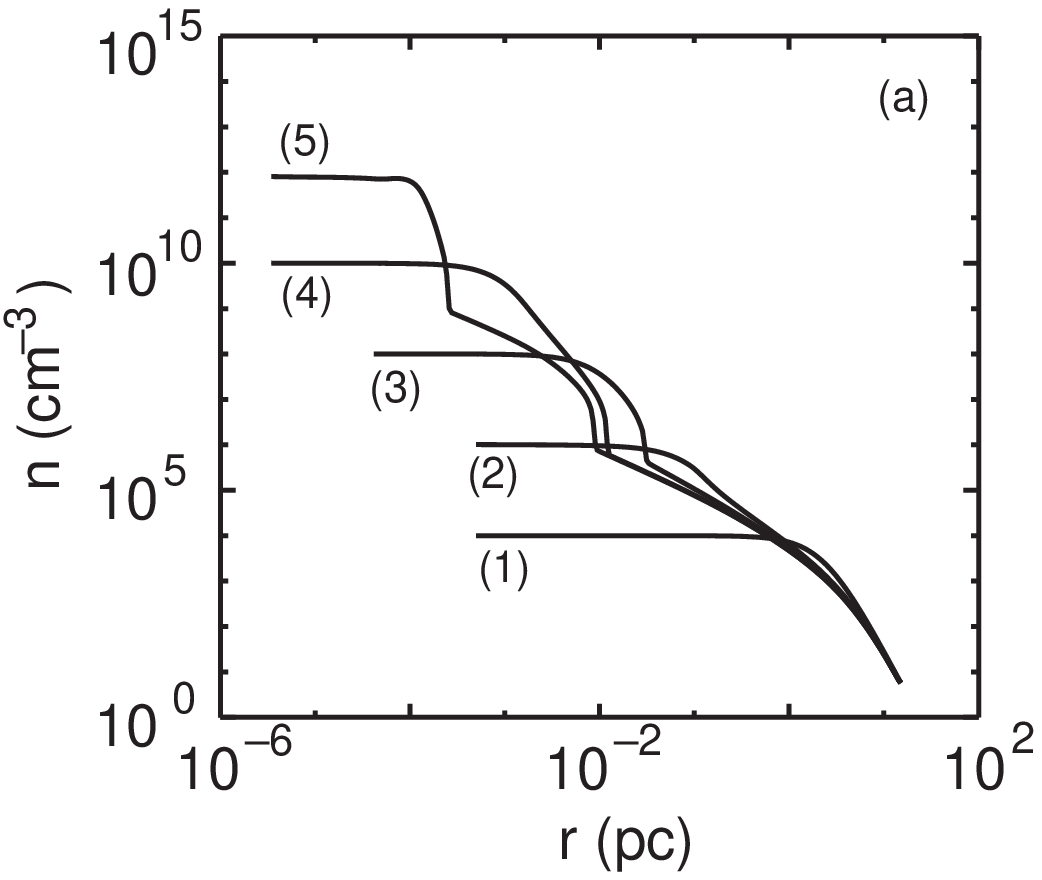}{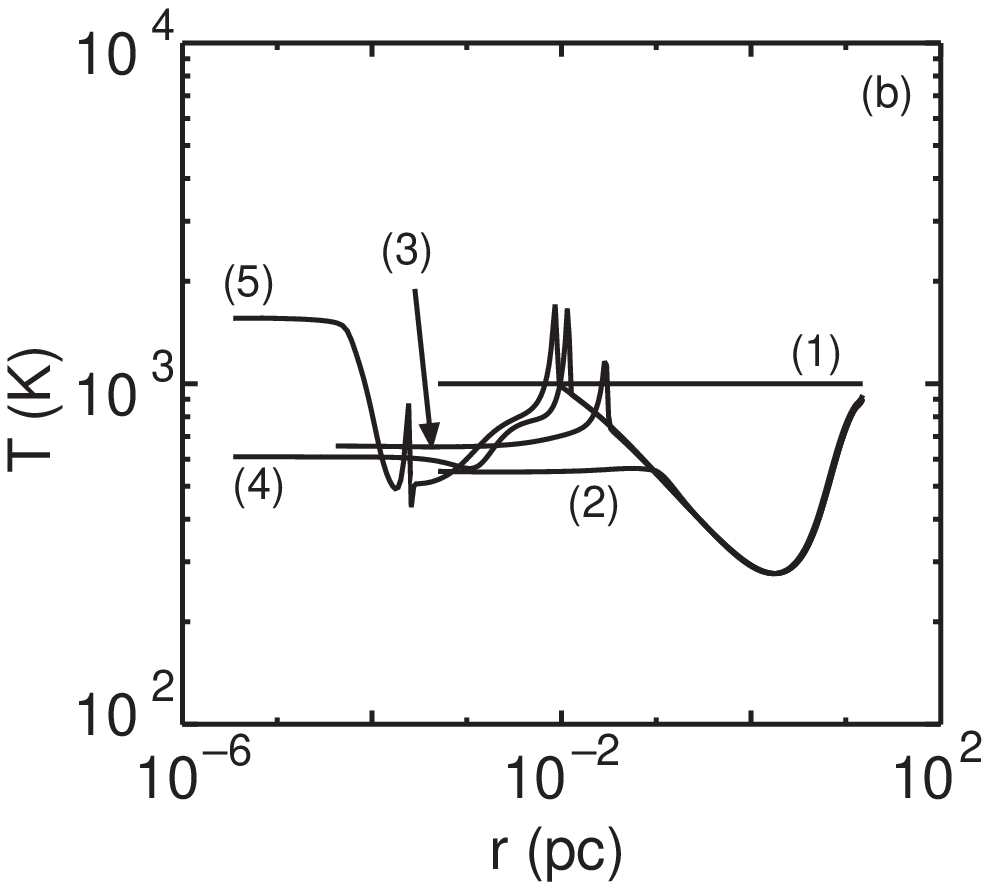} \\
\plottwo{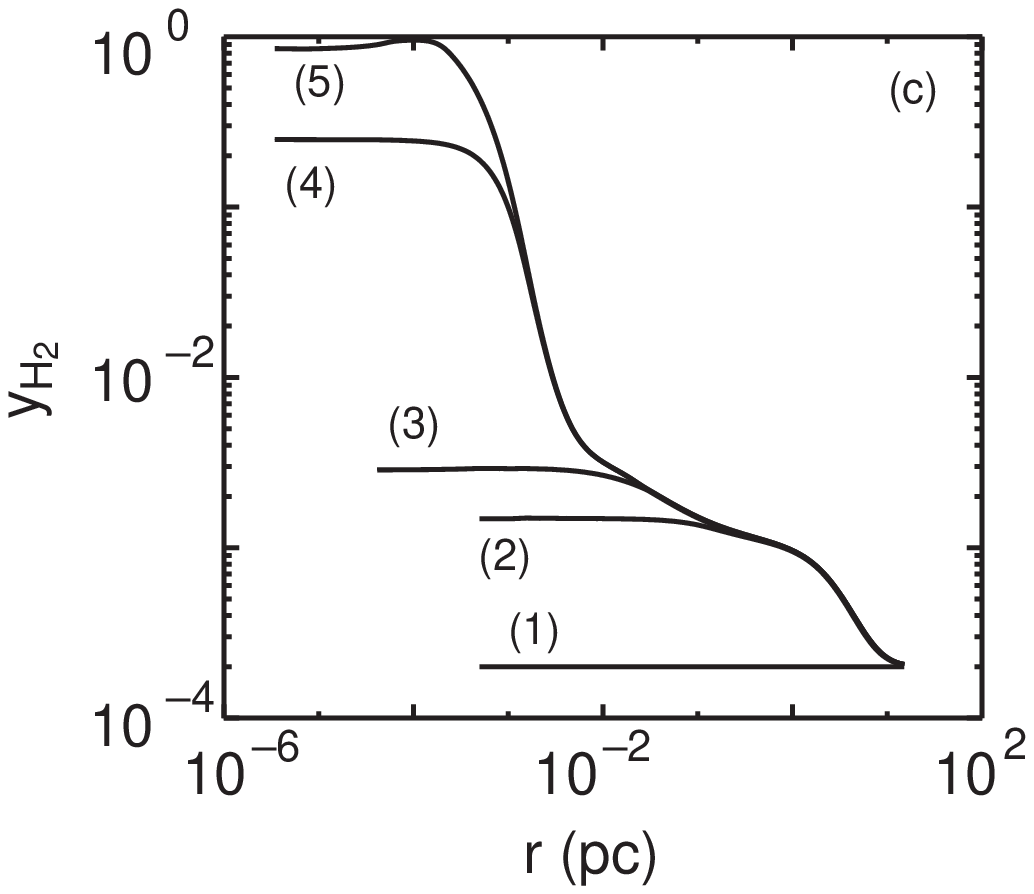}{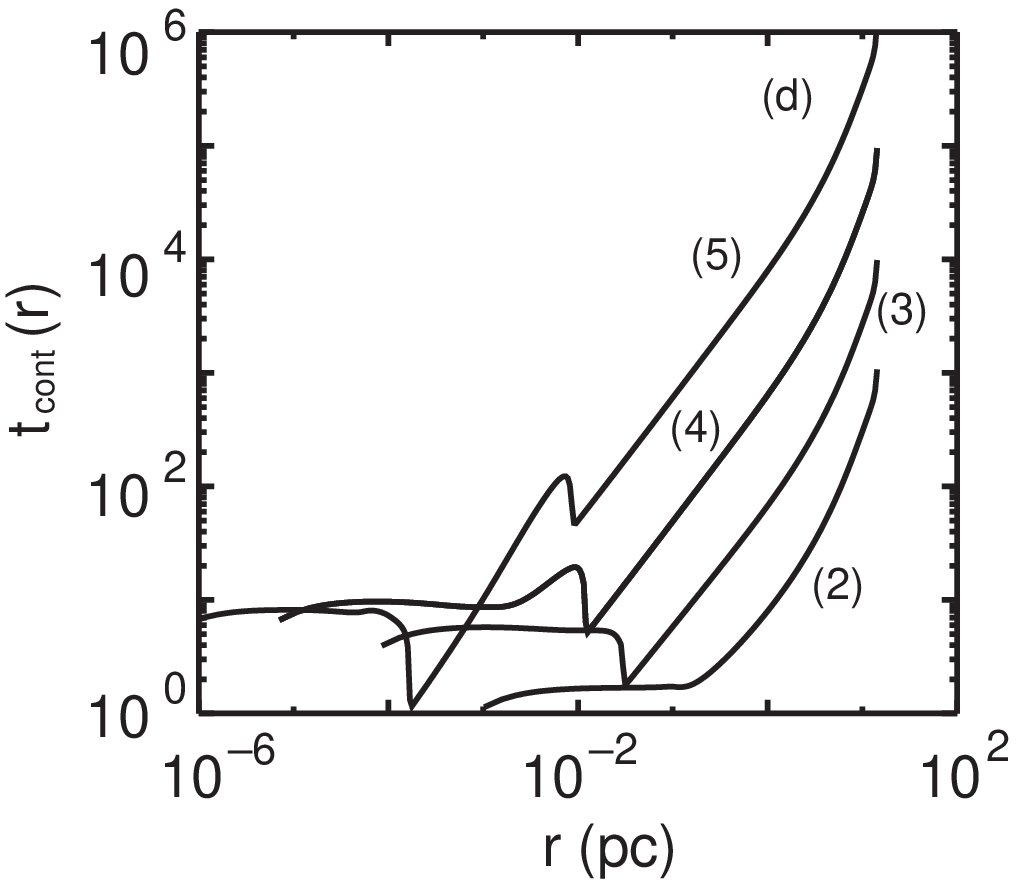}
\caption{Time variations of the distributions for
(a) the density, (b) the temperature, 
(c) the H$_2$ abundance, and (d) the contraction time are shown 
for the same model as Fig. 1, i.e., 
$(n_0, T_0, f)=(10^4 \, {\rm cm}^{-3} ,\ 10^3 {\rm  K}, \ 2)$. 
They are plotted at the stages at which  
(1) $t = 0$ yr ($n_c = 10^4$) cm$^{-3}$, 
(2)  $t = 1.285 \times 10^6$ yr ($n_c = 10^6$ cm$^{-3}$),
(3)  $t = 1.434 \times 10^6$ yr ($n_c = 10^8$ cm$^{-3}$),
(4)  $t = 1.496 \times 10^6$ yr ($n_c = 10^{10}$ cm$^{-3}$), and
(5)  $t = 1.502 \times 10^6$ yr ($n_c = 10^{12}$ cm$^{-3}$).
\label{fig:density etc}}
\end{figure}

\begin{figure}
\plotone{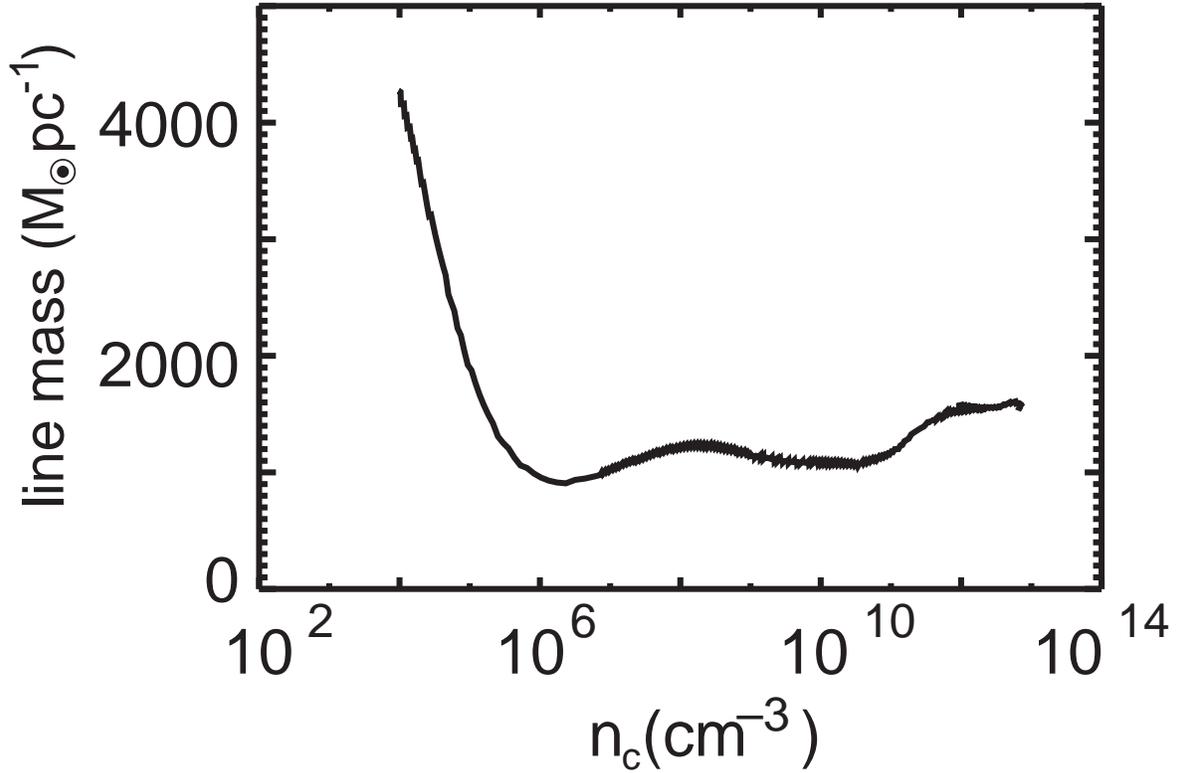}
\caption{Evolution of the line mass of the spindle
in units of $M_\odot$ pc$^{-1}$
for the model with 
$(n_0, T_0, f)=(10^4 \, {\rm cm}^{-3}, \ 10^3 \, {\rm K} , \ 2)$.
The line mass of the spindle keeps a constant 
value, which is $\sim 10^3 M_\odot$ pc$^{-1}$ 
until the cloud becomes optically thick
against the H$_2$ lines. This value accords with 
that of an equilibrium isothermal filamentary cloud with 
the temperature of $T \sim 800 $ K.  It indicates that 
in the spindle the gravitational force nearly balances 
with the pressure force.
\label{fig:line mass}}
\end{figure}

\begin{figure}
\plottwo{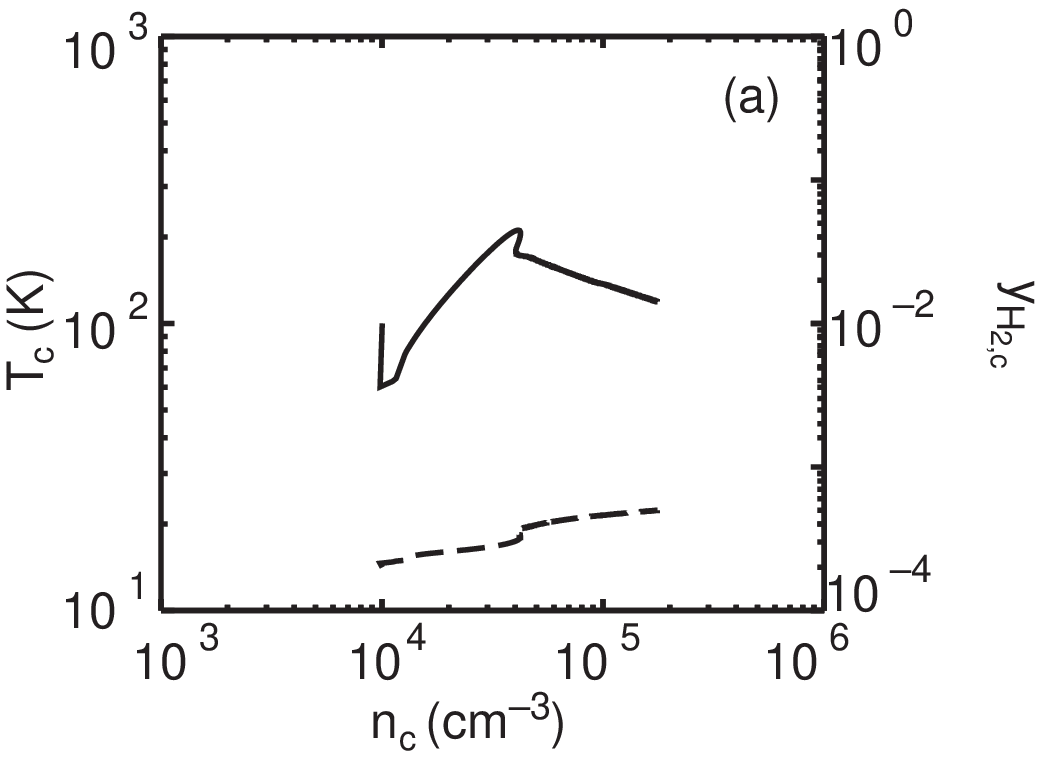}{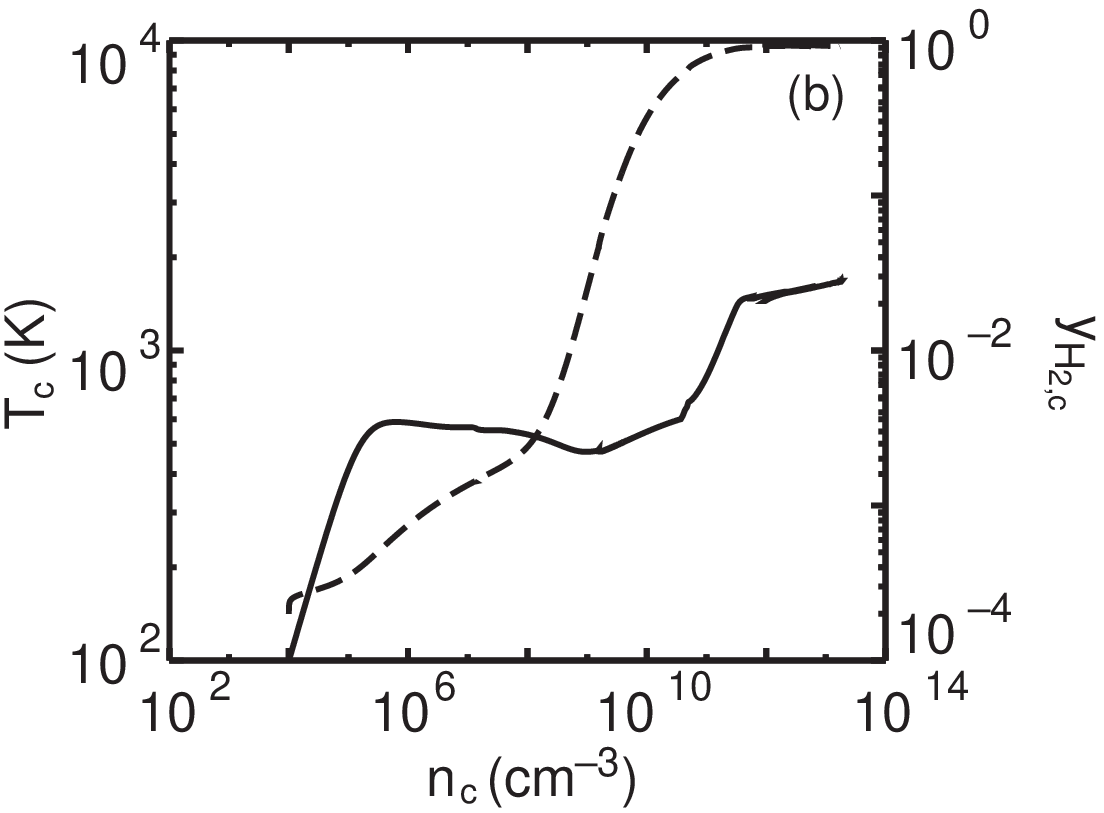}
\caption{Same as Figure \protect\ref{fig:rho-t} but for 
the models with (a) 
$(n_0, T_0, f) = (10^4 \, {\rm cm}^{-3}, 10^2 \, {\rm K}, \ 2)$ and
(b) $(10^4 \, {\rm cm}^{-3}, \ 10^2\, {\rm K}, \ 10)$.
For both models the early evolution is similar; the clouds collapse
adiabatically, $ T \propto \rho^{2/3}$, because 
the H$_2$ line cooling is not effective for such low temperature.
For the model with a small $f$ the collapse almost ceases until the
central density reaches 10$^5$ cm$^{-3}$.
On the other hand, for the model with a large $f$, the evolution is
similar to that of the model with $T_0 \gtrsim 500 $ K.
When the cloud temperature exceeds 500 K, the collapse proceeds nearly 
isothermally due to the H$_2$ line cooling.
\label{fig:low temp}}
\end{figure}

\begin{figure}
\plotone{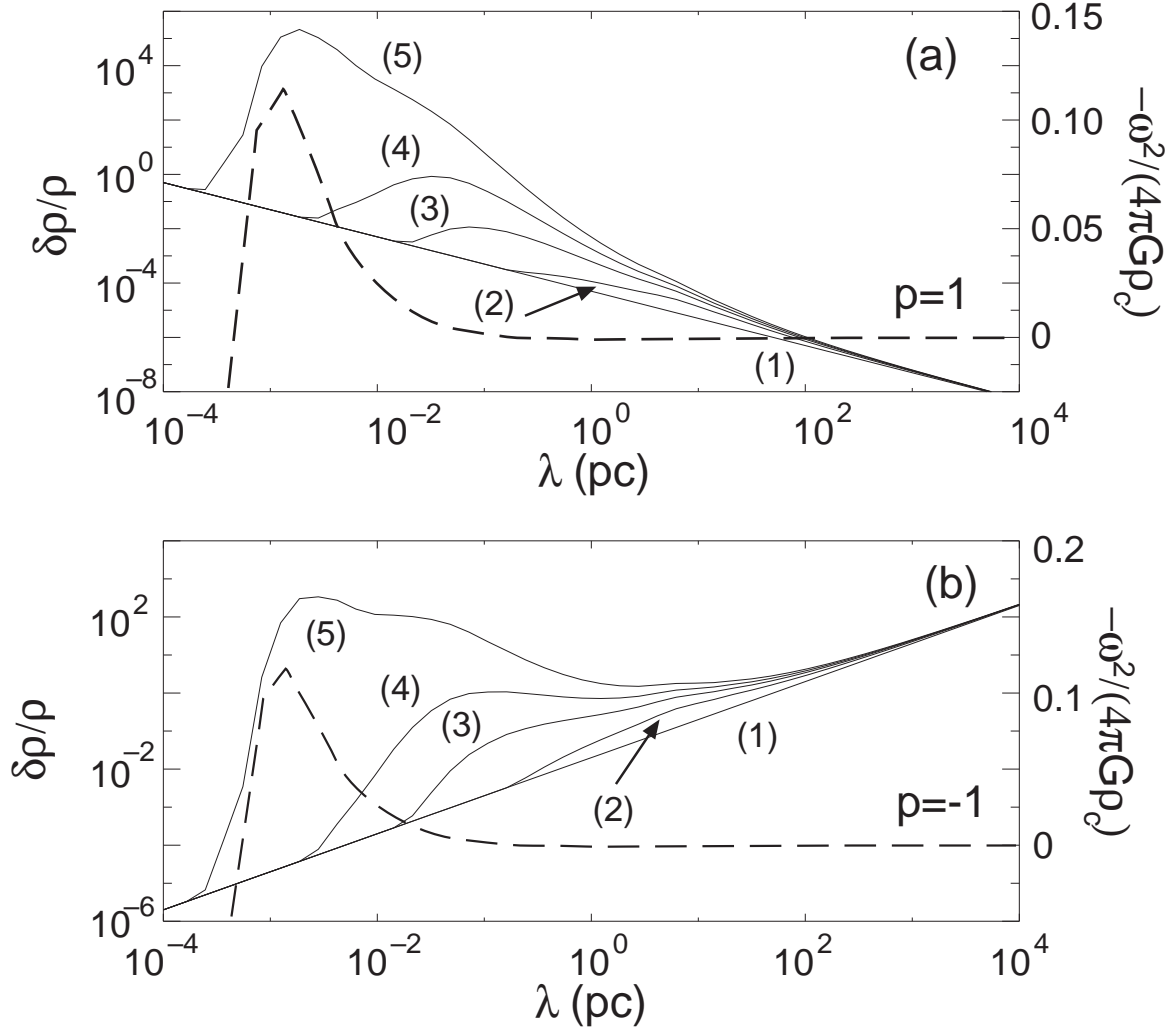}
\caption{Evolution of density perturbations for 
the model with 
$(n_0, T_0, f)=(10^4 \, {\rm cm}^{-3},\ 10^3\, {\rm K},\ 2)$.
The abscissa and ordinate denote the wavelength and amplitude of the
density perturbation, respectively.  
The initial spectrum of the density perturbations 
is assumed to be in the form of $A = A_0 (k_z /k_{z,0})^p$.
The index $p$ is (a) $p = 1$ or (b) $p=-1$. 
We take $(A_0, k_{z,0}) = (1.0 \times 10^{-3},  
1.26 \times 10^2 \, {\rm pc^{-1}})$.
The solid curves represent the amplitude of the density perturbations  
at  (1) $t \, = \, 0$ yr ($n _c \, \simeq \, 10^4 $ cm$^{-3}$), 
(2) $1.384 \times 10^6$ yr (10$^{7}$ cm$^{-3}$), 
(3) $1.477 \times 10^6$ yr (10$^{9}$ cm$^{-3}$), 
(4) $1.500 \times 10^6$ yr (10$^{11}$ cm$^{-3}$), and 
(5) $1.503 \times 10^6$ yr ($3 \times 10^{12}$ cm$^{-3}$),
respectively.
For comparison, we show the square of the 
growth rate at the stage (5) with the dashed curves.
\label{fig:growth}}
\end{figure}

\begin{figure}
\plotone{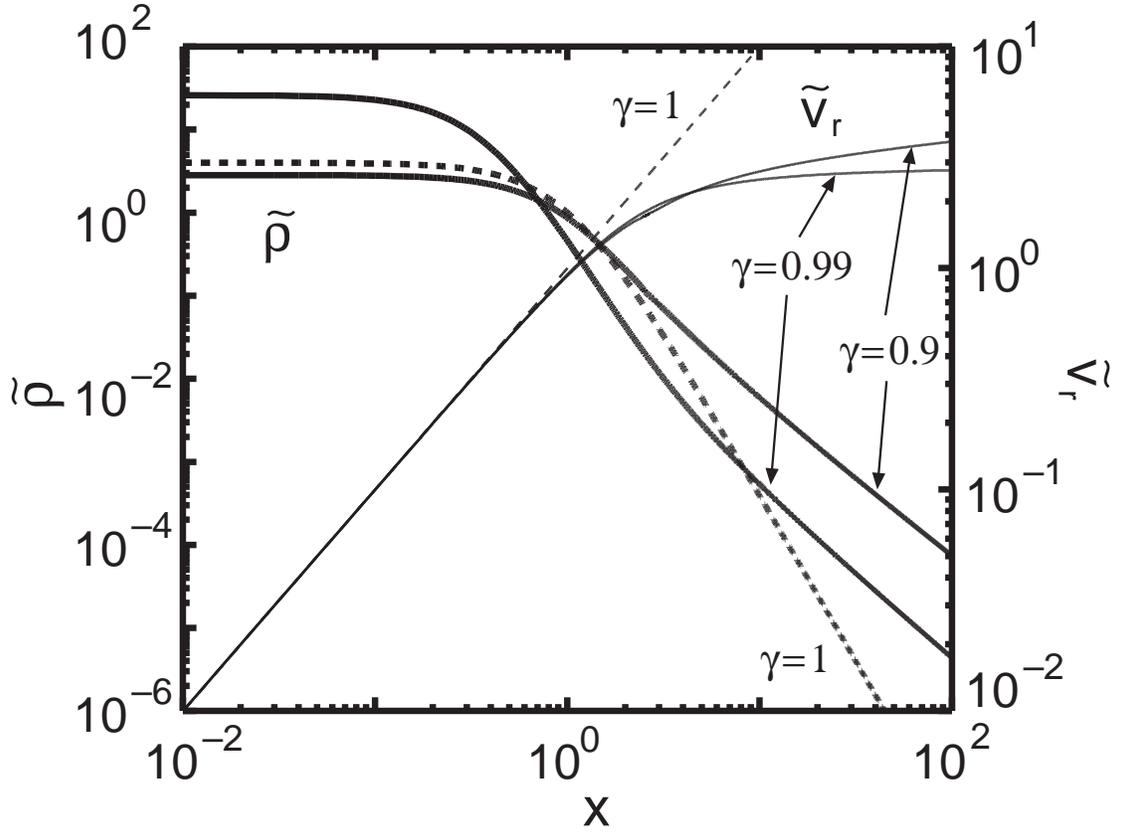}
\caption{Similarity solutions of filamentary clouds with $\gamma = 0.9$
and 0.99.
The thick and thin solid curves show the density and velocity
distributions in the similarity coordinates, respectively.
For comparison, we denote the density and velocity distributions
of the isothermal similarity solution with the thick and thin dashed
curves, respectively.
For the similarity solutions with $\gamma = 0.9$ and 0.99,  
the density distribution is nearly proportional to $r^{-2}$ 
and the velocity converges to a constant value in the outer part.
\label{fig:simil}}
\end{figure}

In Figure \ref{fig:simil} the similarity solutions of 
$\gamma = 0.9 $ and 0.99 are compared with that of $\gamma = 1 $.  
The thick and thin solid curves represent the density and 
velocity distributions of the similarity solutions 
with $\gamma  \ne  1 $, respectively.
The dashed curves are the similarity solution of 
$\gamma  = 1 $ (Miyama et al. 1987a).
The newly found similarity solution 
seems to well reproduce the numerical results 
during the nearly isothermal contraction phase; 
the density distribution in the outer part in the spindle 
is nearly  proportional to $r^{-2}$ and the velocity distribution 
is proportional to $r$ near the center 
[see Fig. \ref{fig:density etc}. Recently Kawachi \& Hanawa (1998)
also studied the gravitational collapse of a polytropic cylinder and
confirmed that the evolution of the cylinder converges to
the similarity solution.].  
It should be stressed again that whether $\gamma$ is precisely unity
or not transforms conclusively the self-similar manner.
The present similarity solution for $\gamma  \ne  1 $ is 
likely to be applicable for the realistic situations.

{}

\clearpage 

\end{document}